\documentclass[aps,prb,onecolumn,superscriptaddress,preprint]{revtex4-1}

\usepackage[utf8]{inputenc}

\usepackage{graphicx}
\usepackage{natbib}
\usepackage{color}
\usepackage[normalem]{ulem}

\newcommand{\tis}{TiS$_2$}
\newcommand{\wse}{WSe$_2$}

\newcommand{\fextis}[1]{Fe$_{#1}$TiS$_2$}

\newcommand{\ncm}[1]{\ensuremath{#1\,\times 10^{20} \textnormal{cm}^{-3}}}
\newcommand{\wkm}[1]{\ensuremath{#1\,\textnormal{W K}^{-1}\textnormal{m}^{-1}}}

\begin{document}
\title{Anisotropic thermal transport in magnetic intercalates \fextis{x}}

\author{Florent Pawula}
\affiliation{Normandie Univ, ENSICAEN, UNICAEN, CNRS, CRISMAT, 14000 Caen, France}

\author{Ramzy~Daou}
\email{ramzy.daou@ensicaen.fr}
\affiliation{Normandie Univ, ENSICAEN, UNICAEN, CNRS, CRISMAT, 14000 Caen, France}

\author{Sylvie~H\'ebert}
\affiliation{Normandie Univ, ENSICAEN, UNICAEN, CNRS, CRISMAT, 14000 Caen, France}

\author{Oleg Lebedev}
\affiliation{Normandie Univ, ENSICAEN, UNICAEN, CNRS, CRISMAT, 14000 Caen, France}

\author{Antoine~Maignan}
\affiliation{Normandie Univ, ENSICAEN, UNICAEN, CNRS, CRISMAT, 14000 Caen, France}

\author{Alaska~Subedi}
\affiliation{Centre de Physique Th\'eorique, \'Ecole Polytechnique, CNRS,
Universit\'e Paris-Saclay, F-91128 Palaiseau, France}
\affiliation{Coll\`ege de France, 11 place Marcelin Berthelot, 75005 Paris, France}

\author{Yohei~Kakefuda}
\affiliation{NIMS, WPI-MANA and CFSN, Tsukuba, Japan}

\author{Naoyuki~Kawamoto}
\affiliation{NIMS, WPI-MANA and CFSN, Tsukuba, Japan}

\author{Tetsuya~Baba}
\affiliation{NIMS, WPI-MANA and CFSN, Tsukuba, Japan}

\author{Takao~Mori}
\affiliation{NIMS, WPI-MANA and CFSN, Tsukuba, Japan}

\date{\today}

\begin{abstract}
We present a study of the of thermal transport in thin single crystals of iron-intercalated titanium disulphide, \fextis{x} for $0\leq x \leq 0.20$.
We determine the distribution of intercalants using high-resolution crystallographic and magnetic measurements, confirming the insertion of Fe without long-range ordering.
We find that iron intercalation perturbs the lattice very little, and suppresses the tendency of \tis{} to self-intercalate with excess Ti. We observe trends in the thermal conductivity that are compatible with our \textit{ab initio} calculations of thermal transport in perfectly stoichiometric \tis{}.
\end{abstract}
\maketitle

\section{Introduction}

As a result of their reduced dimensionality, layered materials can present a wide variety of interesting or useful physical properties. While exotic states of matter can be accessed in the few- or single-layer limit, it is difficult to explore the dimension perpendicular to the plane. Methods of coupling in this direction are often limited to optical or magnetic responses, as transport properties out-of-plane are difficult to access in nanometrically thin materials.

Intercalation of layered compounds can also be used to modify dimensionality and physical properties at the nanoscale, but this can in principle be extended to any thickness. This allows us to engineer and exploit bulk properties in the out-of-plane, as well as the in-plane direction. 
The general principle behind intercalation-tuned properties is either charge transfer from the intercalants to the conduction band; or modification of the inter-layer separation. Although similar effects can sometimes be achieved by chemical substitution, intercalation does not disrupt the crystal structure to the same degree and is often a reversible process. 

While there is a paucity of experimental data, \textit{ab initio} calculations suggest that out-of-plane properties in intercalated materials can be remarkably different from those in the plane \cite{Ong2010,Samanta2014}. 

In practice, single crystals of some of the most commonly intercalated materials -- transition metal dichalcogenides -- are typically grown with a thickness in the van der Waals-coupled direction of some tens of microns, compared to several millimeters laterally. It is quite challenging to measure the anisotropy of electrical and thermal transport properties in such configurations.

\tis{}-based intercalates are well known \cite{Wilson1969} for the sheer variety of possible intercalant species that includes a wide range of transition metals, as well as polar organic molecules. They have been extensively studied as possible thermoelectric materials \cite{Wan2010,Guilmeau2011,Gascoin2012,Beaumale2014} as although the ideal compound is predicted to be a narrow-gap semiconductor, it is difficult to grow with carrier concentration lower than $\sim 10^{20} \textnormal{cm}^{-3}$. This is thought to be the result of self-doping by excess $Ti^{4+}$ intercalated into the van der Waals gap. The weak interlayer coupling and random site intercalation result in reduced in-plane thermal conductivity, while the thermoelectric power is relatively high. Meanwhile, there is some evidence that intercalation by guest species can suppress or mask this tendency to self-intercalation\cite{Daou20bla}.

The degree to which the out-of-plane properties, and thus the suitability for thermoelectric applications, are impacted by intercalation remains relatively unknown. A single report of the out-of-plane thermal conductivity at 300K on \tis{} \cite{Imai2001} suggests a relatively low anisotropy of $\kappa_{ab}/\kappa_{c} \sim 1.5$, where $\kappa_i$ is the thermal conductivity measured in direction $i$.

In this article we present a study of the anisotropic thermal transport, $\kappa_{ab}$ and $\kappa_{c}$, in \fextis{x} for $0<x<0.20$. We use crystallographic and magnetic measurements to confirm the intercalant density and lack of long-range order. We find that iron intercalation perturbs the lattice very little, and suppresses the tendency of \tis{} to self-intercalate with excess Ti. We observe trends in the thermal conductivity that are compatible with our \textit{ab initio} calculations of thermal transport in perfectly stoichiometric \tis{}.

\section{Methods}
Samples of \fextis{x} with nominal concentrations $x=0.0,0.05,0.10,0.20$ were grown by iodine vapour transport. A \tis{} precursor was prepared by reacting pure Ti and S elements at 650$^\circ$C in a closed silica ampoule. \fextis{x} powders were then obtained by reacting stoichiometric amounts of Fe metal and \tis{} under the same conditions.  After synthesis, 1.5g of iodine was mixed with 5g of each \fextis{x} powder, and these powders were sealed under vacuum in 20cm-long closed ampoules. A two-zone horizontal box furnace generated a temperature gradient running from 800$^\circ$C to 900$^\circ$C along the ampoule. After 30 hours, the ampoules were cooled to room temperature in 16 hours. Large platelet-like crystals several millimeters across and some tens of micrometers thick were extracted from the batch.

Room temperature powder X-ray diffraction confirmed that the crystals crystallize in the \tis{} structure (space group: $P\bar{3}m1$), where nominally charge-neutral layers of edge sharing $\mathrm{TiS_6}$ octahedra are separated by a van der Waals gap. 
Since only small changes of the lattice parameters (see Table~\ref{tab:one}) were observed as a function of $x$ compared to other intercalated species \cite{Guilmeau2011,Beaumale2014b,Thompson1975,Li2004,Zhang2011}, additional characterization with transmission electron microscopy (TEM) was used to confirm the presence of intercalants. 
The grids were prepared by crushing crystals in an agate mortar in ethanol, and then depositing the resulting powder on the Cu holey carbon grid. Electron diffraction (ED) studies were performed on an FEI Tecnay G2 LaB6 microscope. High angle angular dark field scanning TEM (HAADF-STEM) images were obtained using a JEM ARM200F cold FEG double-aberration-corrected microscope operated at 200kV and equipped with CENTURIO large-angle EDX detector and QUANTUM GIF.

\begin{figure}
\includegraphics[width=0.7\textwidth]{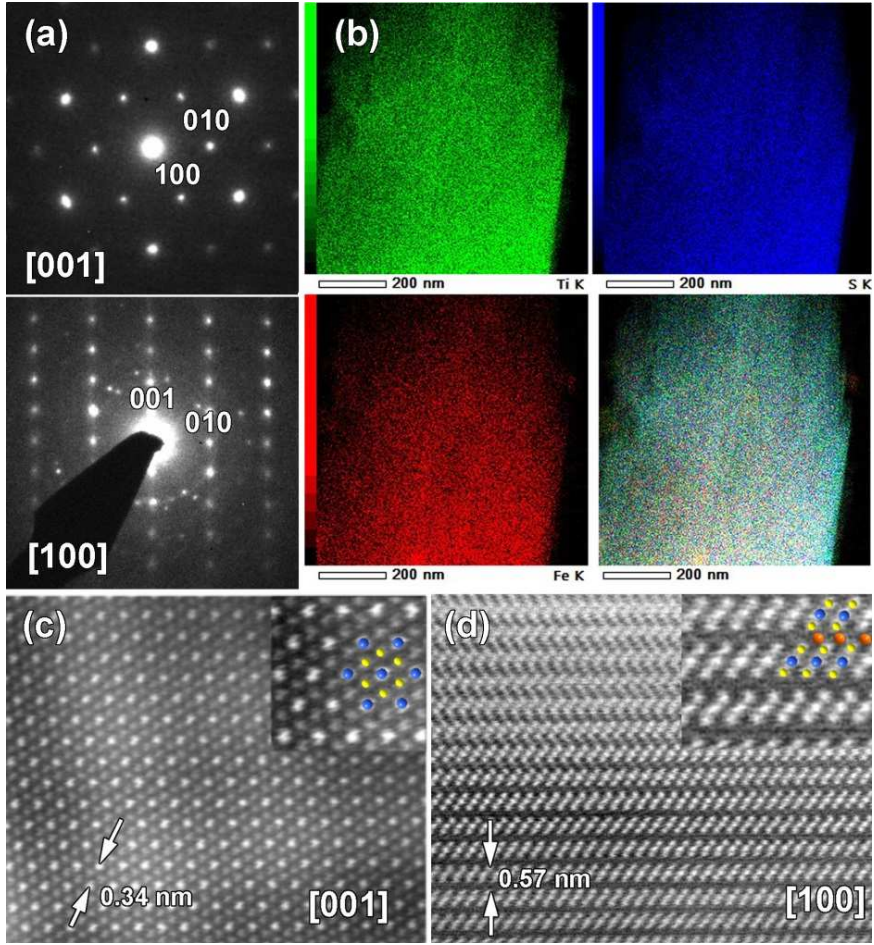}
\caption{ED patterns along the main zone axis – [001] and [100] (b) EDX mapping of elemental Ti K (green) S K (blue) Fe K (red) and overlaid color map of selected crystallite. Note the presence of Fe-rich line defects. (c),(d) – corresponding high resolution HAADF-STEM images of the \fextis{0.20} crystal along [001] and [100] zone axis. The inset shows enlarged images overlaid with the structural model (Ti-blue, S-yellow, Fe-orange).}
 \label{fig:crys}
\end{figure}

Both ED and HAADF-STEM imaging confirm the good crystallinity of the material. This is illustrated by the ED patterns collected for \fextis{0.20} crystals (nominal composition) from the main zone axis (Fig.~\ref{fig:crys}(a)) which are indexed with the same $P\bar{3}m1$ space group. High resolution HAADF-STEM images along [001] and [100] show good correspondence with the basic structure (Fig.~\ref{fig:crys}c and \ref{fig:crys}d) and clear insertion of Fe inbetween \tis{} layers as is shown in the [100] HAADF-STEM image. Local EDX analyses confirmed that the crystals contain Fe in quantities close to the nominal $x$ content. For example, for the $x=0.20$ sample, an average value of $x_{eds}=0.19$ was obtained over 20 measurements. There were, however, a few outliers, which motivated more systematic EDX mapping. This revealed a not completely uniform Fe distribution (Fig.~\ref{fig:crys}b). 

\begin{figure}
 \includegraphics[width=0.7\textwidth]{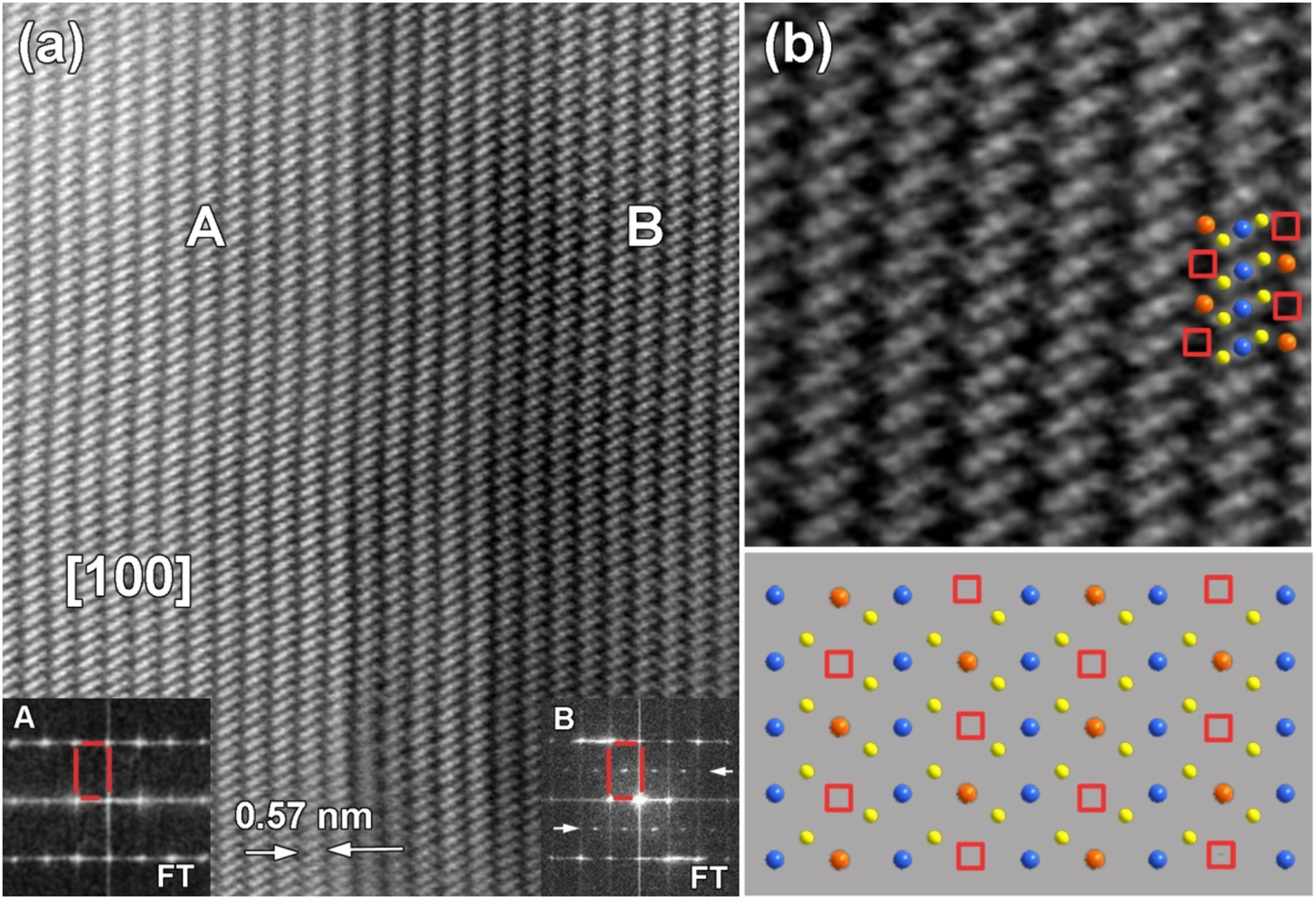}
 \caption{(a) [100] HAADF-STEM image and corresponding FT patterns of intergrowth of two related local structures – A – basic and B – showing superstructure as it is evidence from FT patterns. The superstructure spots denoted with white arrows. (b) – enlargement of superstructure (B) area with superimposed structural model and proposed model of Fe-vacancies ordering (Ti-blue, S-yellow, Fe-orange, vacancy-red square).}
 \label{fig:crys2}
\end{figure}

This is illustrated in Figure~\ref{fig:crys2} where there are two intergrown regions without (A) and with (B) superstructure, as shown in both Fourier Transform (FT) patterns and HAADF-STEM. The HAADF-STEM image indicates that in (A), the Fe intercalated atoms are randomly distributed, while in (B) the superstructure spots come from Fe and vacancies locally ordering in the van der Waals gap (Fig.~\ref{fig:crys2}b). A local model can be superimposed on to the high resolution image. Such intergrowth makes sense, as several extended structures based on similar hexagonal symmetry were reported for the rational filling fractions $x=1/4$, $x=1/3$ and $x=1/2$ in \fextis{x} \cite{Takahashi1973}.

\begin{figure}
 \includegraphics[width=0.7\textwidth]{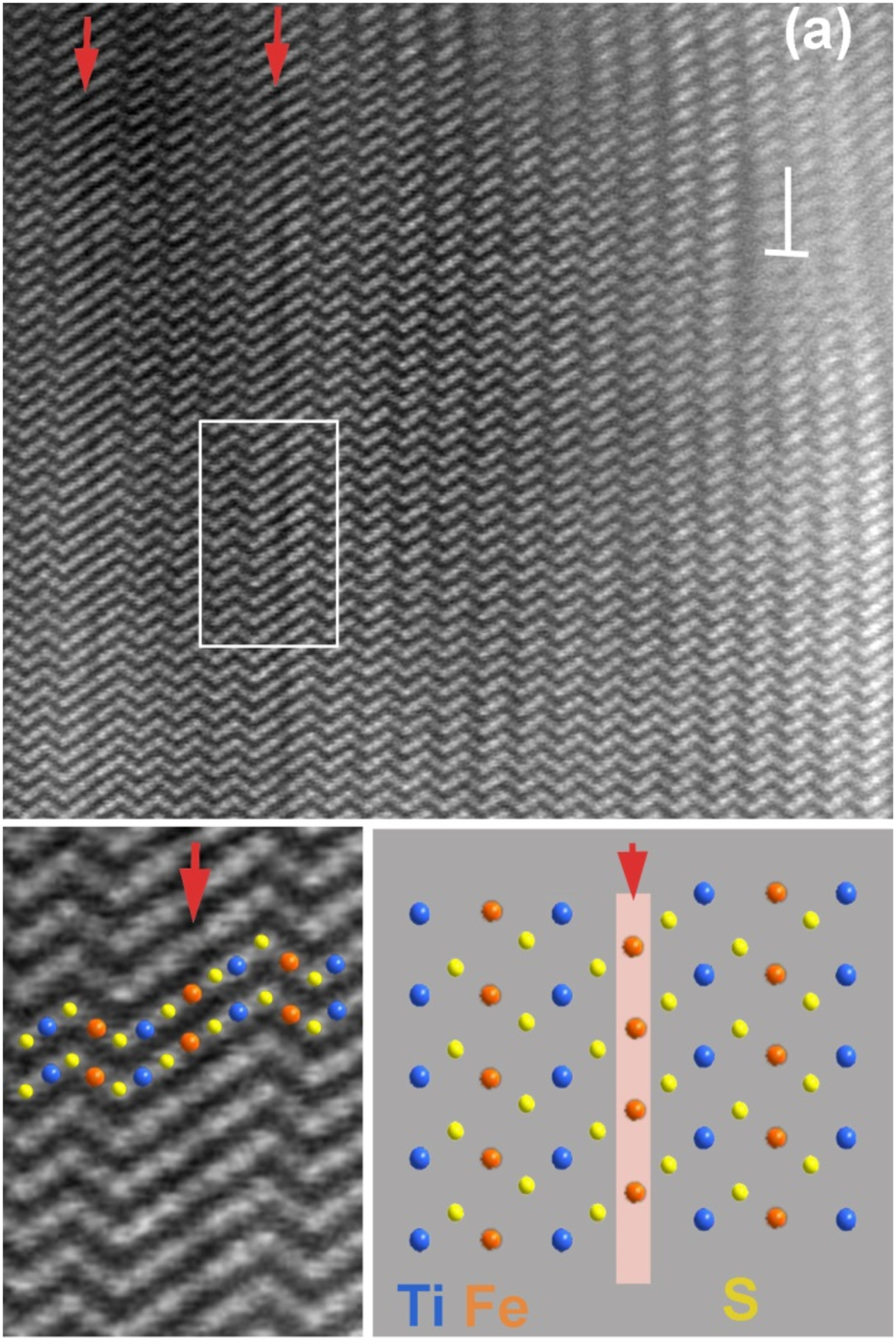}
 \caption{High resolution [100] HAADF-STEM image of a defective area showing stacking faults (marked with red arrows) and misfit dislocations. The enlarged image of a stacking fault (marked with white rectangle) and proposed structural model are shown as insets.}
 \label{fig:crys3}
\end{figure}

Some regions also exhibit planar defects due to inhomogeneous distribution of Fe as shown in Figure~\ref{fig:crys3}. One of the common planar defects is a stacking fault (Fig.~\ref{fig:crys3}a) which is distributed along the c-axis. Such defects were never observed before in pure \tis{} structure. This is probably related to regions with extra Fe atoms that favor the layer shift. Some misfit dislocations, which are normally not so typical for the bulk materials, have also been observed.

Our structural characterizations demonstrate that Fe is inserted in the van der Waals gap in quantities very close to the nominal ones. While there exist some very local defects, related to deviations of the Fe content, which lead to local Fe ordering phenomena or planar defects, there is no evidence of long-range ordering of Fe intercalants.
This is supported by magnetisation measurements, which were performed in the MPMS SQUID magnetometer with a magnetic field of 0.01\,T applied parallel to the crystalline c-axis. 
\fextis{x} shows spin-glass behaviour for $0.01\lesssim<x\lesssim 0.20$ and cluster-spin-glass behaviour for larger values of $x$. The magnetism has a strong Ising character with the easy axis normal to the cleavage plane.
Our results confirm the previously observed spin-glass behaviour for $x=0.05,0.10,0.20$ with gradually increasing glass transition temperatures $T_f = 12,20,30$\,K respectively. The spin-glass phase is another confirmation of the essentially random organisation of the Fe ions; no long-range magnetic order is sustained.

In-plane measurements of transport properties were made on single crystal platelets of typical dimension $5 \times 1 \times 0.02$\,mm. Electrical contacts were made using Dupont 4929 silver paste. Electrical resistivity was measured using a standard four probe technique. In-plane thermal conductivity and thermoelectric power were measured on single crystals using a custom built steady-state experiment installed in a Physical Properties Measurement System cryostat. A calibrated heat pipe was used to measure the thermal power entering the sample \cite{Allen1994}. Power loss (principally from radiation) did not exceed 30\% at 300\,K. This contributes to an additional uncertainty in the thermal conductivity measurements of approximately 10\% at 300\,K, which becomes negligible below 150\,K. 
The systematic error in the results is dominated by the uncertainty in sample thickness (measured by electron microscopy), leading to an uncertainty of approximately 20\%.

The cross-plane thermal conductivity was also explored using picosecond time-domain thermoreflectance (TD-TR) instruments (PicoTherm Corp., PicoTR). Cleaved samples were fixed to a flat substrate and a 100nm thick layer of platinum was deposited on their top surface to detect transient surface temperature changes through the related change in reflectivity. A 1550nm infrared pulsed laser light (repetition frequency 20MHz, pulse duration 0.5ps) was used as a heat source. A 780nm probe laser was used to detect surface temperature change. The picosecond TD-TR system was customized to reduce the spot size of the probe laser to $\sim5\mu$m \cite{Kakefuda2017}. By a combination of focused laser light and a near infrared CCD camera, we were able to perform site-selective detection of TD-TR signals from terraced regions of the specimens.

We note that out-of-plane electrical transport measurements (resistivity and thermoelectric power) on our single crystals are not practical, as they are too thin. There is to our knowledge only a single report\cite{Imai2001} of $c$-axis resistivity on a 100\,$\mu$m thick single crystal \tis{}, which has $\rho_{c}(300K) = 1.3\Omega$\,cm,  $\sim 750\times\rho_{ab}(300K)$. This level of anisotropy is similar to other layered van-der-Waals materials such as graphite.

The thermal conductivity of un-intercalated, defect-free \tis{} was calculated from first
principles taking into account the three-phonon scattering
processes. This involved first calculating the phonon dispersions,
which were obtained using density functional perturbation theory as
implemented in the {\sc quantum espresso} package \cite{dfpt,qe}. We
used pseudopotentials generated by Garrity \textit{et al.}
\cite{gbrv} and planewave expansions with cutoffs of 50 and 250 Ryd
for the basis set and charge density, respectively. The Brillouin zone
integration was performed using a $16\times16\times8$ grid. The
dynamical matrices were calculated on a $8\times8\times4$
grid. Grimme's semiempirical recipe was used to treat the van der
Waals contribution to the potential energy \cite{grimme}.

Calculating the three-phonon scattering processes requires computing
the third-order interatomic force constants. These were obtained from
the supercell-based finite difference method using the
\texttt{thirdorder.py} code \cite{shengbte}. We used $4\times4\times2$
supercells and a force cutoff distance of third nearest neighbors in
these calculations. The phonon contribution to the lattice thermal
conductivity was then calculated by iteratively solving the Boltzmann
transport equation on a $24\times24\times12$ grid using the {\sc
  shengbte} package \cite{shengbte}. All calculations were performed
using the fully-relaxed structure with lattice parameters $a = b =
3.387$ and $c = 5.824$ \AA\ and atomic positions Ti (0,0,0) and S
$(1/3,2/3,0.245)$. 


\section{Results}

\begin{figure}
 \includegraphics{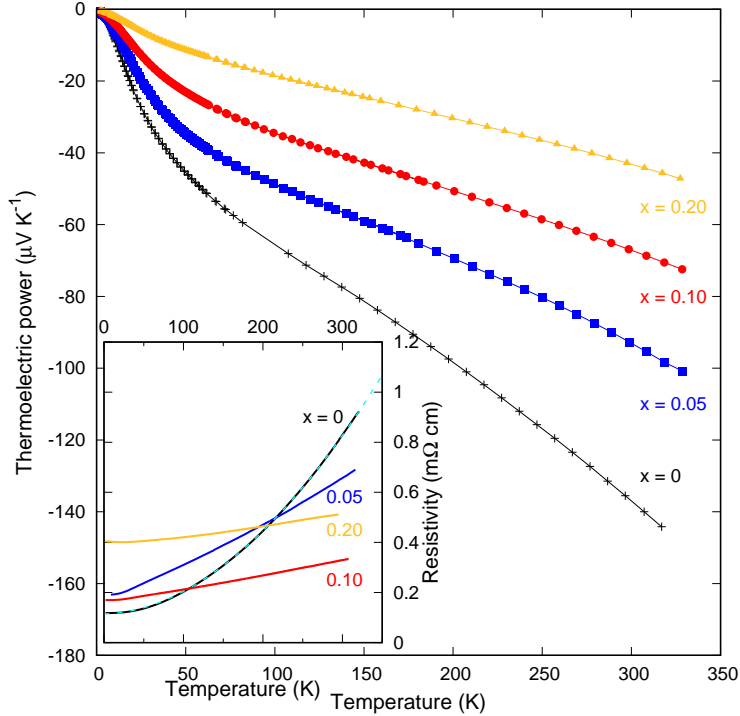}
\caption{The thermoelectric power of \fextis{x}. A monotonic reduction of the thermoelectric power with $x$ indicates a steady increase of the carrier concentration. Inset: Measured in-plane resistivity of \fextis{x}. The temperature dependence is reduced as $x$ increases. The fit to the resistivity at $x=0$ (see text for details) is shown as a dashed cyan line.}
\label{fig:transport}
\end{figure}

The thermoelectric power $S$ and electrical resistivity $\rho$ of the four samples is shown in Figure~\ref{fig:transport}. The low-temperature resistivity increases and becomes less dependent on temperature as the level of Fe-intercalation increases. The evolution is not monotonic with $x$, however. While intercalation adds carriers, it also introduces scattering centers. Interpreting the resistivity is also complicated by systematic errors, in particular uncertainty in the sample geometry. The evolution of the thermoelectric power is more regular, implying a steady increase in carrier concentration with $x$.

The room temperature Hall coefficient $R_H$ of \tis{} yields a carrier concentration $n_H = \ncm{5.1}$, consistent with a self-intercalation level $y=0.007$ in Ti$_{1+y}$S$_2$, assuming that each excess intercalated Ti$^{4+}$ contributes 4$e^-$ to the conduction band. $y=0.008$ is obtained by comparison of $S(300K)=-136\mu$V K$^{-1}$ to the sample series in Ref.~\onlinecite{Thompson1975}. The resistivity at $x=0$ can be fit to a form $\rho = \rho_0 + AT^\alpha$ with $\alpha=1.9$ over the range 20--300\,K. $\alpha$ is also known to evolve with carrier concentration, and this value is consistent with this Hall coefficient\cite{Klipstein1981}.

Because of the strong anomalous contribution arising from the magnetic intercalants, $R_H$ cannot be used to determine $n_H$ for the Fe-intercalated samples. In this case we determine the carrier effective mass $m^*$ from the single parabolic band (SPB) approximation for doped semiconductors, using $n_H(x=0)$ and $S(300K)$ \cite{May2009}. We assume that the obtained value $m^* = 4.8m_e$ does not change with $x$, and then use the values of $S(300K)$ for $x>0$ to estimate the carrier concentrations, $n_{SPB}$. The relevant equations are reproduced in Appendix~\ref{sec:appendix}.

The carrier concentrations are closely comparable to the nominal intercalant density $n_x$, which would suggest that every intercalated Fe donates one electron to the conduction band. The agreement between $n_x$ and $n_{H,SPB}$ becomes better at high $x$, implying that the introduction of Fe intercalants suppresses either the tendency to or the effect of self-intercalation by excess Ti. These values are summarised in Table~\ref{tab:one}. A tendency for self-intercalation to be incompatible with other forms of intercalation was observed before for intercalated NH$_3$ intercalants \cite{Thompson1975}.

\begin{table*}[t]
 \begin{tabular}{ c | c c c  c  c  c  c  c c}
  \hline
  x   & a & c & $n_x$                            & $n_H$ &  $n_{SPB}$                          & $\rho_{0}$     & $S_{ab}$  &
  $\kappa_{ab}$ & $\kappa_{c}$ \\
      & (\AA) & (\AA) & ($10^{20} \textnormal{cm}^{-3}$) & ($10^{20} \textnormal{cm}^{-3}$) & ($10^{20} \textnormal{cm}^{-3}$) &(m$\Omega$cm) & ($\mu$VK$^{-1}$) &
   (WK$^{-1}$m$^{-1}$)  & (WK$^{-1}$m$^{-1}$)  \\
  \hline
  0.00 & 3.400(1) & 5.701(1) & 0.0 & 5.1 &  & 0.12 & -136 & 3.8(0.9) & 0.63(0.36) \\
  0.05 & 3.409(1) & 5.704(1) & 8.7 &  & 10.5 & 0.19 & -93 & 5.1(1.3) & 0.92(0.20)  \\
  0.10 & 3.410(1) & 5.702(1) & 17.4 &  & 17.7 & 0.17 & -67 & 6.6(1.6) & 0.99(0.30) \\
  0.20 & 3.410(1) & 5.702(1) & 34.8 &  & 34.8 & 0.40 & -43 & 3.4(0.8) & 1.06(0.40) \\
  \hline
 \end{tabular}
\caption{Transport properties of \fextis{x} at 300K. The nominal dopant density $n_x$ is compared to the carrier density $n_H$ derived from the Hall coefficient (for the $x=0$ sample) and $n_{SPB}$ from the single parabolic band approximation ($x>0$) via the thermoelectric power at 300\,K.}
\label{tab:one}
\end{table*}

\begin{figure}
 \includegraphics{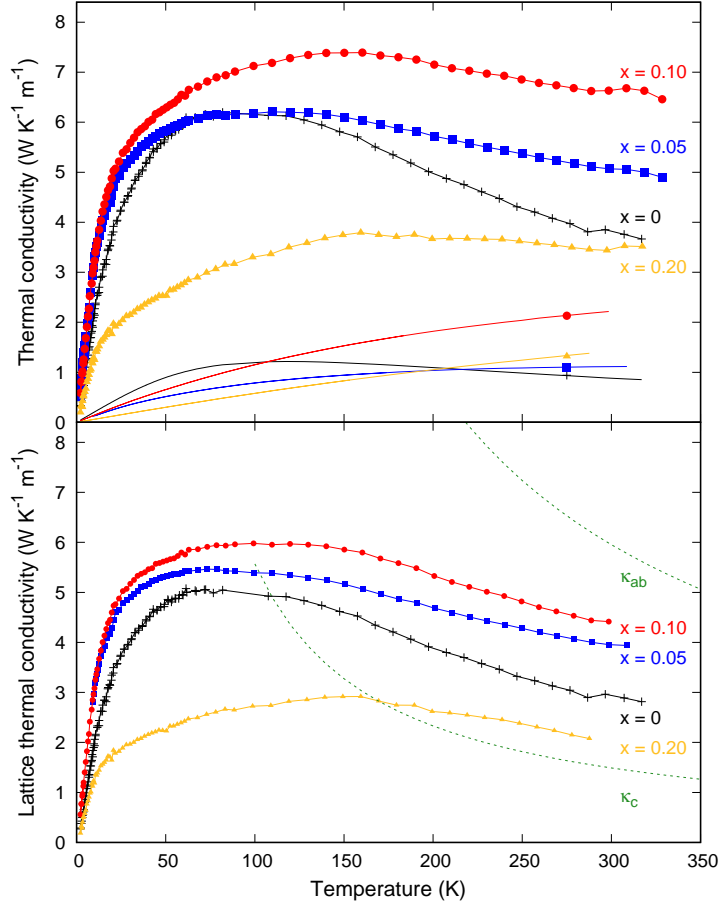}
\caption{Top: In-plane thermal conductivity of \fextis{x}. At 300\,K, There is little change from $x=0$ to $x=0.10$ within the uncertainty of the measurement. Thermal conductivity is strongly suppressed for $x=0.20$. The electronic contribution to the thermal conductivity (solid lines) is derived from the resistivity data using the Wiedemann-Franz law, $\kappa_e = LT/\rho$, where $\rho(T)$ is taken from Fig.~\ref{fig:transport}. The symbols and colors indicate to which measured thermal conductivity curve these lines correspond. Bottom: the in-plane lattice contribution to the thermal conductivity resulting from the subtraction of the electronic contribution. The green dashed curves show the calculated lattice thermal conductivities in the in-plane ($\kappa_{ab}$) and out-of-plane ($\kappa_{c}$) directions. }
\label{fig:thermal}
\end{figure}

The in-plane thermal conductivity data are shown in Figure~\ref{fig:thermal}. The form is typical of nominally stoichiometric \tis{} and lightly doped derivatives, with a broad peak at around 80\,K that becomes flattened with increasing $x$. While the data suggests that the thermal conductivity initially increases with $x$, this increase is mostly within the uncertainty arising from the sample geometry. The dominant uncertainty is in the measurement of sample thickness, which is difficult to establish to within $\pm 20\%$. For $x=0.20$ it is clear that the thermal conductivity is strongly reduced, as has been seen in other heavily intercalated \tis{} compounds \cite{Daou20bla}.

The lattice contribution to the in-plane thermal conductivity can be obtained by subtracting the electronic contribution, $\kappa^e$ (Figure~\ref{fig:thermal}). This can be estimated using the Wiedemann-Franz law: $\kappa^e_{ab} = LT/\rho_{ab}$, where $L=2.44\times \textnormal{10}^{-8}\textnormal{W} \Omega\textnormal{K}^{-2}$, where $\rho_{ab}(T)$ is obtained from Fig.~\ref{fig:transport}.

\begin{figure}
 \includegraphics[width=0.48\textwidth]{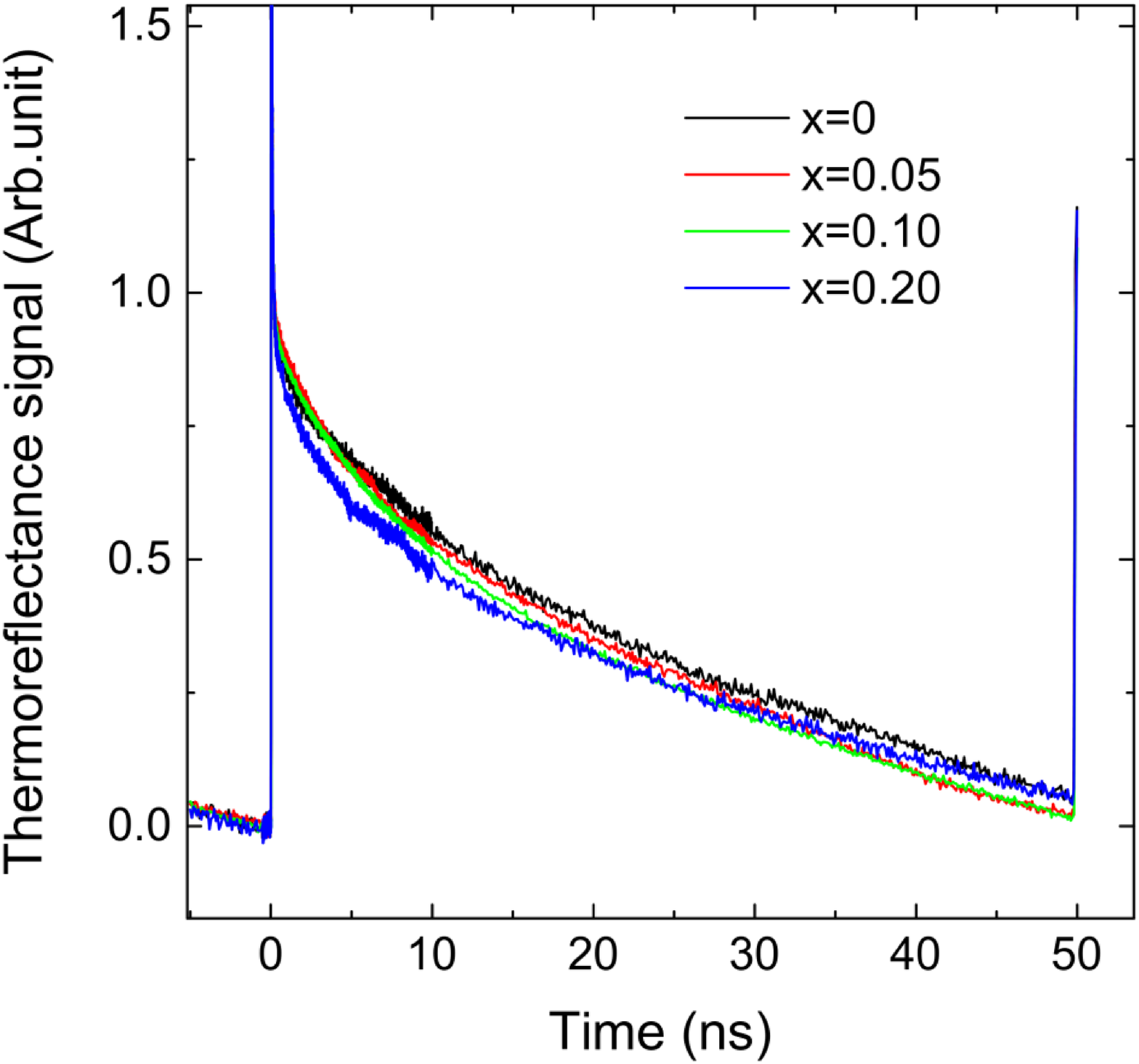}
 
 \includegraphics[width=0.48\textwidth]{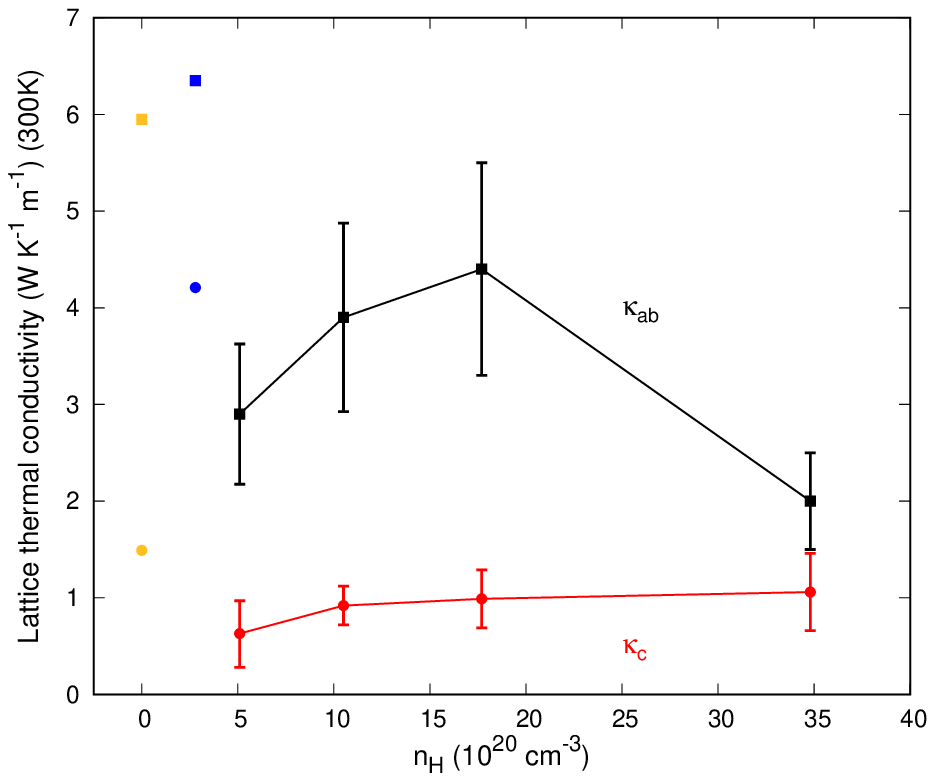}
\caption{Top: Thermoreflectance signals from \fextis{x} samples at 300\,K. An analytical model of the experiment is used to extract the out-of-plane thermal conductivity by fitting the curves from the peak of the signal up to 50\,ns. For more details see Refs.~\onlinecite{Baba2009, Baba2018}. Bottom: The lattice thermal conductivities $\kappa_{c}$ (red circles) and $\kappa_{ab}$ (black squares) at 300K as a function of the carrier concentration $n_H$. The calculated results at $x=0$ are also shown (yellow circle/square), as well as the data point of Ref.~\onlinecite{Imai2001} (blue circle/square). The electronic contribution to $\kappa_c$ is assumed to be negligible.}
\label{fig:thermoreflectance}
\end{figure}

Since the samples grow as thin platelets, it is difficult to access the out-of-plane thermal response with standard techniques. A single data point for the c-axis thermal conductivity has been reported. At 300\,K, $\kappa_c = \wkm{4.2}$ was obtained from a laser-flash diffusivity measurement \cite{Imai2001}. This technique is difficult to apply to thin samples and the results may not be reliable. Time-domain thermoreflectance (TDTR) is a fast optical technique that is well adapted to measuring the cross-plane thermal conductivity of thin platelet crystals\cite{Cahill2004,Chiritescu2007}.
A series of the thermoreflectance signals is shown in Figure~\ref{fig:thermoreflectance}. The cross-plane thermal conductivity values were extracted by using the mirror image method \cite{Baba2009,Baba2018}. This method of analysis is independent of the absolute intensity of the signal.

We estimate that the electronic contribution to the out-of-plane thermal conductivity, $\kappa_c^e$ is negligible. The reported value\cite{Imai2001} of $\rho_{c}(300K) = 1.3\Omega$\,cm for nominally stoichimetric samples with a carrier concentration \ncm{2.8} corresponds to an electronic thermal conductivity, $\kappa_c^e$, of only \wkm{0.0006}, well within the uncertainty of the measurement. 

Our $x=0$ sample has only $\sim 2\times$ this carrier concentration, and we do not anticipate a significant reduction in electrical anisotropy. Therefore we feel justified in assuming that there is also negligible electronic contribution to the thermal conductivity, in this sample at least.

Furthermore, even if the $c$-axis resistivity of the Fe-intercalated samples were to be reduced by $100\times$ relative to this value due to the increased carrier concentration (which seems unlikely, since in our most intercalated sample the carrier concentration is only $\sim 10\times$ greater than Ref.~\onlinecite{Imai2001}), the electronic contribution to the thermal conductivity would still be small enough to neglect. The $c$-axis resistivity would have to grow as large as the in-plane resistivity for the $x>0$ samples in order to account for the trend seen in $\kappa_c$ as a function of $x$. This seems unlikely in materials that retain a strong structural anisotropy and van der Waals gap.

We therefore feel justified in assuming that $\kappa_c^e$ is negligible for all samples discussed here, and that the out-of-plane thermal conductivity arises almost exclusively from lattice vibrations.

\section{Discussion}

$\kappa_{c}$ obtained from the thermoreflectance measurements is considerably lower than the previously reported value for \tis{}, but is close to the value observed in similarly-structured \wse{} \cite{Chiritescu2007}. This implies a strong lattice anisotropy. This is not unexpected in layered van der Waals-coupled materials. In pyrolitic graphite, the lattice thermal anisotropy can be of the order of 50-100 \cite{Slack1962}, for example. The rather high value of $\kappa_c$ reported by the laser flash technique \cite{Imai2001} (blue symbols in Fig~\ref{fig:thermoreflectance}) may therefore be viewed as anomalous in this respect. Dense textured ceramic samples prepared by spark plasma sintering also present a degree of thermal anisotropy\cite{Guilmeau2017}, although it is difficult to extrapolate from such results to the single crystal case due to the limited degree of orientation of the platelets and the unknown contribution of grain boundaries.

Interestingly, the behavior of the intercalation $x$-dependence of $\kappa_c$ is different from that of $\kappa_{ab}$. We found that $\kappa_c$ increases modestly as the amount of intercalated Fe is increased. Because the c-axis lattice parameter does not appear to change significantly, phonon conduction in this direction should not be depressed due to the enhancement of interlayer distance. We suggest that as intercalation density increases, the intercalated Fe atoms may start to function as phonon conduction paths rather than as phonon scattering sources, at least in this direction. This is consistent with the HAADF-STEM result shown in Fig.~\ref{fig:crys2}b, which reveals the existence of some local interaction between Fe atoms.

These phonon conduction paths would not serve to facilitate transport in the plane, so within this scenario $\kappa_{ab}$ would be expected to decline monotonically over the entire doping range. This is consistent with the in-plane data. We note that since the lattice parameter does not appear to change with $x$, it would seem that this replacement induces little distortion in the lattice. Other intercalants, including excess Ti\cite{Beaumale2014b}, have a greater effect on the lattice parameters at much lower concentrations.

The calculated lattice thermal conductivites for both in-plane and out-of-plane directions on \tis{} are shown in Figure~\ref{fig:thermal}. $\kappa_{ab}$ is considerably higher in the calculation than in the measurement at 300\,K, and this trend becomes more significant at low temperatures. This is further evidence that in-plane thermal transport in self-intercalated \tis{} suffers from considerable scattering arising from intercalants and defects.

The calculated out-of-plane thermal transport falls in the range of the measurement at 300\,K. We might speculate that thermal transport in this direction is already limited by scattering at the van der Waals gap, and that the presence of a modest number of additional scattering centers in the gap would not cause a significant change in $\kappa_c$. The relative insensitivity of $\kappa_c$ to additional Fe intercalants supports this argument. This argument is not incompatible with the suggestion that Fe intercalants can act as phonon conduction paths in modest concentrations. As the scattering by intercalants is reduced, a minimum scattering rate due to the van der Waals gap is maintained, leading to an upper limit on $\kappa_c$.

\section{Conclusions}
Stoichiometric defect-free \tis{} does not appear to exist outside of \textit{ab initio} calculations.
The contemporary hypothesis is that excess Ti$^{4+}$ intercalates into the van der Waals gap. We have argued that Fe intercalation competes against this process, and that each Fe contributes only one electron to the conduction band. 
Additionally, crystallographic measurements show that Fe intercalation perturbs the lattice to a much lesser degree than other intercalants. 
Perhaps counter-intuitively, taken together these observations imply that for larger $x$ the \tis{} layers in \fextis{x} are more representative of the hypothetical pristine compound, albeit with a significant number of charge carriers inserted. 

The evolution of $\kappa_{ab}$ and $\kappa_c$ as a function of $x$ are consistent with the suppression of self-intercalation. While the presence of more Fe in the van der Waals gap appears to modestly enhance $\kappa_c$ at the same time as reducing $\kappa_{ab}$, these trends are extended in $x$. Most clearly, the relative insensitivity of $\kappa_c$ to $x$ indicates that the low value obtained from \textit{ab initio} calculations is appropriate. 

A clear trend in $\kappa_{ab}$ is obscured by the experimental value for our $x=0$ sample, which is somewhat lower than the value of $\kappa_{ab}$ from Ref.~\onlinecite{Imai2001}, for a single crystal with half the carrier concentration. The value there is much closer to the one calculated, suggesting a greater sensitivity of $\kappa_{ab}$ to $y$ in Ti$_{1+y}$S$_2$. It would be of interest to study the interplay between different intercalating species, including excess Ti, more systematically. These results and previous studies\cite{Daou20bla} suggest that the effect is not purely additive. Co-intercalation of different species in a controlled way could shed light on this issue.

\begin{acknowledgments}
This work was supported by the Swiss National Supercomputing Center (CSCS) under project s575.
\end{acknowledgments}

\appendix
\section{Single Parabolic Band Approximation}
\label{sec:appendix}
The single parabolic band approximation (SPB) is discussed in detail in Ref.~\onlinecite{May2009}. The validity of the SPB and the calculations based on it are that the transport can be represented by the Boltzmann transport equations, within the relaxation time approximation. Further, electronic conduction occurs within a single parabolic band, and a single scattering mechanism dominates the transport properties, where the energy dependence of the carrier relaxation time follows a simple power law.

These conditions are a reasonable approximation for \tis{}. A single band crosses the Fermi energy resulting in six equivalent Fermi surface pockets. The Fermi energy is small, less than 0.2~eV above the bottom of this band for the highest carrier concentration. While the relevant band is anisotropic, for transport purposes it can be approximated by an isotropic band with an isotropic effective mass.

Here we reproduce the equations relevant to the determination of the carrier concentration. The thermoelectric power is the simplest quantity:
\begin{equation}
 S = \frac{k}{e}
 \left ( 
 \frac{(2 + \lambda)F_{1+\lambda}(\eta)}{(1 + \lambda)F_{\lambda}(\eta)} - \eta
 \right )
\end{equation}
where $k$ is Boltzmann's constant, $e$ is the electronic charge, $\eta = \varepsilon_F e/kT$ is the reduced Fermi energy (or chemical potential), $\varepsilon_F$ is the Fermi energy and
\begin{equation}
 F_i = \int_0^\infty \frac{\xi^i d\xi}{1+\exp(\xi-\eta)}
\end{equation}
is a Fermi integral. $\lambda$ is a parameter that depends on the type of scattering. We have assumed that acoustic phonon scattering dominates, thus $\lambda=0$. By adjusting $\eta$ we can reproduce the values of $S$ for each sample at $T=300$K.

The Hall carrier density is then given by:
\begin{equation}
 n_{SPB} = 4\pi \left ( \frac{2m^*kT}{h^2} \right ) \frac{F_{1/2}(\eta)}{r_H}
\end{equation}
where $m^*$ is the effective mass of the carriers, and $h$ is Planck's constant. $r_H$ is a factor that is close to unity for our samples that is given by:
\begin{equation}
 r_H = \frac{3}{2} F_{1/2}(\eta) \frac{(1/2 + 2\lambda)F_{2\lambda-1/2}(\eta)}{(1 + \lambda)^2F_{\lambda}^2(\eta)}
\end{equation}

$m^*$ was chosen such that the calculated value of $n_{SPB}$ agreed with the measured value $n_H$ at $x=0$. This value of $m^*$ was assumed to be the same for all values of $x$.

\end{document}